\documentclass[prl,superscriptaddress,amsmath,amssymb,twocolumn]{revtex4}
\usepackage{natbib}
\usepackage{graphicx}         
\usepackage{dcolumn}          
\usepackage{bm}               
\usepackage{upgreek}
\usepackage{booktabs} 				
\usepackage{tabularx}
\usepackage{array}
\usepackage[colorlinks=true,urlcolor=blue,breaklinks=true]{hyperref}
\usepackage{xcolor}

\begin{document}

\title{Phase-coherent loops in selectively-grown topological insulator nanoribbons}

\author{J. K\"olzer}
\affiliation{Peter Gr\"unberg Institut (PGI-9) and JARA-Fundamentals of Future Information Technology, J\"ulich-Aachen Research Alliance, Forschungszentrum J\"ulich, 52425 J\"ulich, Germany}

\author{D. Rosenbach}
\affiliation{Peter Gr\"unberg Institut (PGI-9) and JARA-Fundamentals of Future Information Technology, J\"ulich-Aachen Research Alliance, Forschungszentrum J\"ulich, 52425 J\"ulich, Germany}

\author{C. Weyrich}
\affiliation{Peter Gr\"unberg Institut (PGI-9) and JARA-Fundamentals of Future Information Technology, J\"ulich-Aachen Research Alliance, Forschungszentrum J\"ulich, 52425 J\"ulich, Germany}

\author{T. W. Schmitt}
\affiliation{Peter Gr\"unberg Institut (PGI-9) and JARA-Fundamentals of Future Information Technology, J\"ulich-Aachen Research Alliance, Forschungszentrum J\"ulich, 52425 J\"ulich, Germany}
\affiliation{JARA-FIT Institute Green IT, RWTH Aachen University, 52056 Aachen, Germany}

\author{M. Schleenvoigt}
\affiliation{Peter Gr\"unberg Institut (PGI-9) and JARA-Fundamentals of Future Information Technology, J\"ulich-Aachen Research Alliance, Forschungszentrum J\"ulich, 52425 J\"ulich, Germany}

\author{A. R. Jalil}
\affiliation{Peter Gr\"unberg Institut (PGI-9) and JARA-Fundamentals of Future Information Technology, J\"ulich-Aachen Research Alliance, Forschungszentrum J\"ulich, 52425 J\"ulich, Germany}

\author{P. Sch\"uffelgen}
\affiliation{Peter Gr\"unberg Institut (PGI-9) and JARA-Fundamentals of Future Information Technology, J\"ulich-Aachen Research Alliance, Forschungszentrum J\"ulich, 52425 J\"ulich, Germany}

\author{G. Mussler}
\affiliation{Peter Gr\"unberg Institut (PGI-9) and JARA-Fundamentals of Future Information Technology, J\"ulich-Aachen Research Alliance, Forschungszentrum J\"ulich, 52425 J\"ulich, Germany}

\author{V. E. Sacksteder IV}
\affiliation{Wittenberg University, Springfield, Ohio, 45504, United States}
\affiliation{Royal  Holloway  University  of  London, Egham  Hill,  Egham,  TW20  0EX,  United  Kingdom}

\author{D. Gr\"utzmacher}
\affiliation{Peter Gr\"unberg Institut (PGI-9) and JARA-Fundamentals of Future Information Technology, J\"ulich-Aachen Research Alliance, Forschungszentrum J\"ulich, 52425 J\"ulich, Germany}

\author{H. L\"uth}
\affiliation{Peter Gr\"unberg Institut (PGI-9) and JARA-Fundamentals of Future Information Technology, J\"ulich-Aachen Research Alliance, Forschungszentrum J\"ulich, 52425 J\"ulich, Germany}

\author{Th.~Sch\"apers}
\email{th.schaepers@fz-juelich.de}
\affiliation{Peter Gr\"unberg Institut (PGI-9) and JARA-Fundamentals of Future Information Technology, J\"ulich-Aachen Research Alliance, Forschungszentrum J\"ulich, 52425 J\"ulich, Germany}

\hyphenation{}
\date{\today}

\begin{abstract}
Universal conductance fluctuations and the weak antilocalization effect are defect structure specific fingerprints in the magnetoconductance that are caused by electron interference. Experimental evidence is presented that the conductance fluctuations in the present topological insulator (Bi$_{0.57}$Sb$_{0.43}$)$_2$Te$_3$ nanoribbons which are selectively grown by molecular beam epitaxy are caused by well-defined and sharply resolved phase-coherent loops. From measurements at different magnetic field tilt angles we deduced that these loops are preferentially oriented parallel to the quintuple layers of the topological insulator material. Both from a theoretical analysis of universal conductance fluctuations and from weak antilocalization measured at low temperature the electronic phase-coherence lengths $l_\phi$ are extracted, which is found to be larger in the former case. Possible reasons for this deviation are discussed.  
\end{abstract}
\maketitle

\section{Introduction}

Nanoribbons of topological insulators (TI) have attracted considerable interest recently, in particular, since in combination with superconducting contacts spatially separated Majorana excitations are expected to be realized for the purpose of preparing robust topological quantum bits \cite{Cook11,Manousakis17}. In this context the understanding of quantum transport in such TI nanoribbons is of paramount importance, particularly in connection with the topologically protected surface states \cite{Bardarson10,Zhang10,Sacksteder16}. So far most studies in this direction have been performed on ribbons which have been grown epitaxially in a bottom-up approach from the gaseous phase \cite{Peng10,Yan13a,Hamdou13a,Hong14,Jauregui16,Arango16,Dufouleur17}. In such ribbons well-developed Aharonov--Bohm (AB) oscillations due to magneto-transport within the topologically protected surface states occurred \cite{Peng10,Xiu11,Tian13,Hamdou13a,Hong14,Cho15,Jauregui16,Arango16,Dufouleur17,Ziegler18}. The two-dimensional nature of the surface states in nanoribbons was confirmed by observing Shubnikov--de Haas oscillations \cite{Xiu11,Fang12,Gooth14}, while phase-coherent transport was revealed by the presence of conductance fluctuations and weak antilocalization \cite{Matsuo12,Matsuo13,Arango16,Dufouleur17}. 

With respect to more circuit flexibility in networks of nanoribbons for topological quantum computation a planar arrangement of these structures prepared by lithography in a top-down process would be advantageous, as compared to the standard bottom-up approach of preparing single free standing nanoribbons. In order to achieve this goal, we fabricated planar (Bi$_{0.57}$Sb$_{0.43}$)$_2$Te$_3$  nanoribbons which are grown selectively by molecular beam epitaxy (MBE) within lithographically patterned nano-grooves. This particular stoichiometry was chosen in order to minimize the bulk conductivity contribution \cite{Weyrich16}. In the selectively grown nanoribbons the van der Waals bonded quintuple layers are oriented parallel to the nanoribbon axis. Thus, current transport between contacts at both ends of the ribbon occurs parallel to the quintuple layers. 

For designing circuits for topological quantum computation, it is important to know on which length scale phase coherence is maintained. This information can be obtained by analyzing phase-coherent transport phenomena such as weak (anti)localization and universal conductance fluctuations \cite{Beenakker91c}. In both cases the magnetotransport is governed by electron interference in an ensemble of closed-loop trajectories of different size. The loops are formed by elastic scattering at randomly distributed impurities. In the case of weak localization, the constructive interference of time-reversed paths results in an enhanced resistance at zero magnetic field. By applying a magnetic field $\vec{B}$ a distinct phase-shift between these time-reversed paths occurs, which depends on the encircled magnetic flux $\Phi$. Since the cross-section with respect to the magnetic field is different for each loop, different phase shifts are accumulated so that the constructive interference is lost. As a result, the average resistance decreases towards the classical value upon increasing $B$. Interference changes sign when spin scattering is present or spin precession occurs, resulting in weak antilocalization rather than weak localization and a resistance dip rather than a resistance peak at zero field \cite{Hikami80,Bergmann84}. 

For nano-scaled disordered samples with just a finite number of scattering centers only a few closed-loops $\Lambda$ are present. Since all of these loops have a different cross section $\vec{S}_\Lambda$ with respect to $\vec{B}$, each particular loop $\Lambda$ acquires a different phase shift due to the different encircled flux $\Phi_\Lambda$. According to the Aharonov--Bohm effect the conductance contribution of each closed loop will be periodic with the magnetic flux quantum $\Phi_0=h/e$: $G(\Phi_\Lambda) = G(\Phi_\Lambda + \Phi_0)$. Since each loop has a different cross-section $\vec{S}_\Lambda$ and thus encircles a different flux the corresponding magnetic field period $\Delta B_\Lambda$ is also different for each loop. As a result, the superposition of oscillations with different periods leads to a fluctuating magneto-conductance pattern  \cite{Lee85,Altshuler85b}. The fluctuation pattern is individual for each sample but reproducible when repeating the measurement. Vice versa, distinct features observed in the frequency spectrum of the magnetoconductance can be directly assigned to closed loops with a well-determined cross section $\vec{S}_\Lambda$ with respect to a given magnetic field orientation \cite{Sacksteder18}. 
By performing transport measurements at different tilt angles of the magnetic field, information on the spatial orientation of the phase-coherent loops can be gained \cite{Jespersen15}. Following the conceptual framework of phase-coherent transport as outlined above, we analyze the magnetotransport of a selectively-grown (Bi$_{0.57}$Sb$_{0.43}$)$_2$Te$_3$ Hall bar as well as a nanoribbon structure to gain information on the scale at which phase-coherence is maintained.   

\section{Experimental} 

\begin{figure}[htb]
	\centering
\includegraphics[width=0.98\columnwidth]{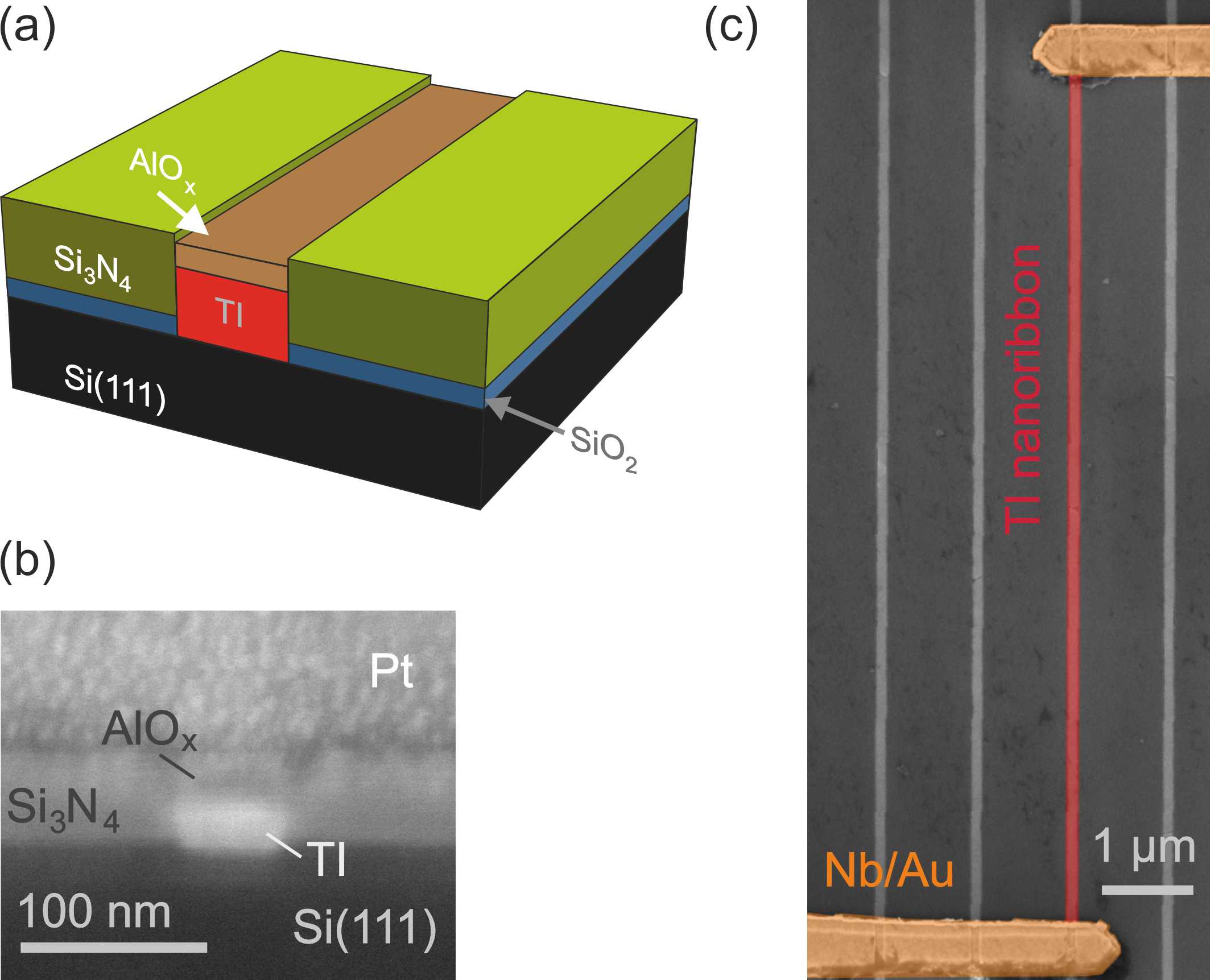}
	\caption{(a) Schematic illustration of a selectively-grown topological insulator (TI) nanoribbon. The nanoribbon is capped by a thin aluminum oxide layer. (b) Focused ion beam cut cross section of a 50-nm-wide nanoribbon. The top Pt layer is deposited to prepare the lamella for electron microscopy. (c) Scanning electron beam micrograph of a contacted 100-nm-wide nanoribbon (indicated in red). The neighboring nanoribbons do not contribute to the transport.}
	\label{fig:wire-TI-SEM-FIB}
\end{figure}

Hall bars and nanostructures have been fabricated by MBE using selective-area growth \cite{Kampmeier16,Weyrich19}. In  Fig.~\ref{fig:wire-TI-SEM-FIB}(a) a schematics of the sample layout is shown. For substrate preparation, first, a silicon wafer with a (111) surface orientation is covered with a 5.8-nm-thick thermally-grown SiO$_{2}$ layer. Subsequently, a 25-nm-thick amorphous Si$_{3}$N$_{4}$ layer is deposited on top using low pressure chemical vapour deposition. A lithographic mask allows to etch the desired structures by reactive ion etching (CHF$_3$/O$_2$) and a hydrofluoric acid wet etching step. Using this process a growth mask on the silicon wafer with smooth crystalline surfaces in the etched areas is achieved. The 29-nm-thick topological insulator film was grown selectively by means of MBE on the Si(111) surface at a temperature of 370$^\circ$C. The temperature window  for selective area growth within the mask opening, where no deposition on the mask surface takes place is remarkably small, i.e. of $365 ^\circ \mathrm{C} < T < 375^\circ \mathrm{C} $. The growth parameters were adjusted such that the stoichiometric composition of the material (Bi$_{0.57}$Sb$_{0.43}$)$_2$Te$_3$ is grown \cite{Weyrich16}. An in-situ capping was achieved by depositing $2\,\mathrm{nm}$ of Al that oxidized once exposed to air \cite{Lang11}. A cross section of a 50-nm-wide nanoribbon prepared by focused ion beam  is shown in Fig.~\ref{fig:wire-TI-SEM-FIB}(b). 

The Ohmic contacts composed of a 5-nm-thick Nb layer and a 100-nm-thick Au layer were sputtered on top of the TI after removing the AlO$_x$ capping in the contact areas by wet chemical etching and argon sputtering. Hall bars were prepared for characterization of the (Bi$_{0.57}$Sb$_{0.43}$)$_2$Te$_3$ film, while the phase-coherent transport in nanostructures was investigated in nanoribbons. A scanning electron micrograph of a contacted nanoribbon fabricated by selective area growth is shown in Fig.~\ref{fig:wire-TI-SEM-FIB}(c).

For magnetotransport measurements a variable temperature insert equipped with a 14\,T magnet was employed. The samples were measured at temperatures between 1.4 and 35\,K. In order to vary the magnetic field angle a rotating sample holder was employed. The electrical measurements were performed by using a lock-in technique and injecting an AC current of $10\,\mathrm{nA}$ into the nanoribbon. 

\section{Results and Discussion}

\subsection{Hall bar structures}

\begin{figure}[htb]
	\centering
\includegraphics[width=0.98\columnwidth]{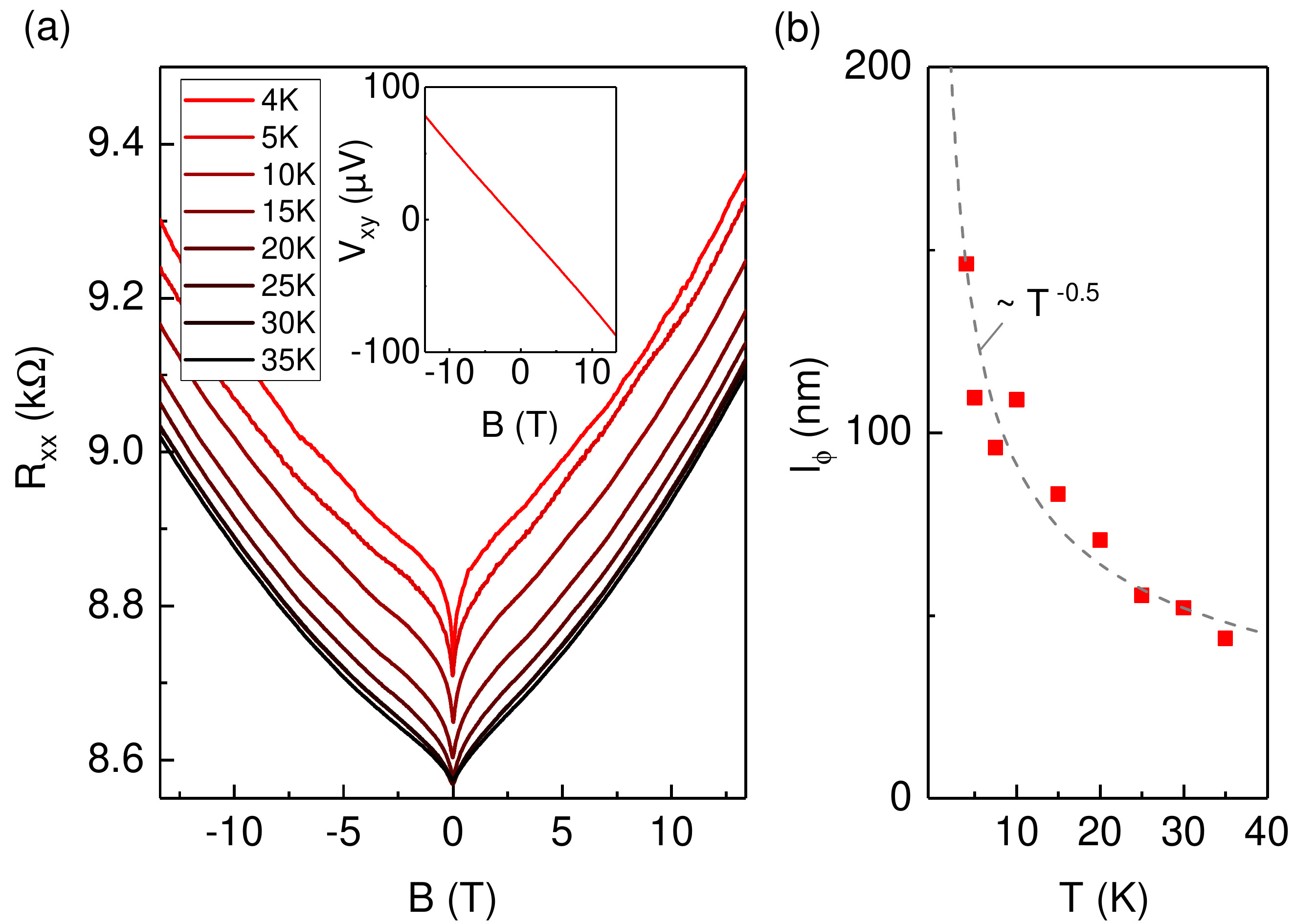}
	\caption{(a) Magnetoconductance of a Hall bar structure measured at temperatures ranging from 4 to 35\,K. The inset shows the corresponding Hall voltage $V_\mathrm{xy}$ at 4\,K. (b) Phase-coherence length $l_\phi$ as a function of temperature extracted from the weak antilocalization measurements. The gray line corresponds to the fit with a temperature dependence according to $l_\phi \propto T^{-0.5}$. $l_\phi$ was obtained by performing a HLN fit to the data.}
	\label{fig:WAL-Hall-bar}
\end{figure}

In order to characterize the transport properties of the (Bi$_{0.57}$Sb$_{0.43}$)$_2$Te$_3$ film, first, magnetoresistance and Hall effect measurements were performed on a Hall bar structure with a length of $100\,\mathrm{\mu m}$ and a width of $10\,\mathrm{\mu m}$. In Fig.~\ref{fig:WAL-Hall-bar}(a) the magnetoresistance at temperatures ranging from 4 to 35\,K is shown. 
The magnetic field is oriented perpendicular to the TI layer. The magnetoconductance curves reveal a positive magnetoresistance with a distinct dip at zero magnetic field, which can be attributed to weak antilocalization \cite{Hikami80,Bergmann84,Adroguer15}, as it was reported for Bi$_2$Se$_3$ \cite{Chen10,Chen11,Steinberg11}, Bi$_2$Te$_3$ \cite{He11,Chiu13,Weyrich16}, Sb$_2$Te$_3$ \cite{Takagaki12a,Taskin12a}, as well as for (Bi$_{x}$Sb$_{1-x}$)$_2$Te$_3$ layers \cite{Weyrich16}. As can be seen in Fig.~\ref{fig:WAL-Hall-bar}(a), inset, the Hall voltage has a linear negative slope, which indicates that the transport is $n$-type \cite{Weyrich16}. From magnetoresistance and Hall measurements a resistivity of $\rho=2.5 \times 10^{-3}\,\Omega$cm, a projected 2D carrier concentration of $1.02 \times 10^{14}\,\mathrm{cm}^{-2}$, and a mobility of $139\,\mathrm{{cm}^{2}/{Vs}}$ at a temperature $T=4\,\mathrm{K}$ were extracted. The relatively large charge carrier concentration indicates a considerable contribution of bulk charge carriers participating in the transport \cite{Weyrich16}. From the above values an elastic mean free path of $l_e=6$\,nm was derived. By increasing the temperature from 4 to 35\,K, the electron concentration only increases slightly, by less than 3\%, showing that the transport is in the metallic regime. In order to gain information on phase-coherent transport from the weak antilocalization feature a fit to the Hikami--Larkin--Nagaoka (HLN) model was performed \cite{Hikami80}. From the fit a phase-coherence length  $l_{\phi}$ of about 150\,nm was extracted at 4\,K. The temperature dependent decay was found to be proportional to $T^{-0.5}$ which is in agreement with the Nyquist electron-electron interaction for disordered systems \cite{Altshuler81a}. Furthermore, we find that the transport takes place in a single channel, since the fitted pre-factor $\alpha$ of the HLN formula of about $-0.4$ is close to the predicted value of $-0.5$ for a single channel. This suggests an interpretation of two topological two-dimensional transport channels that are coupled via bulk scattering \cite{Ando13,Weyrich16}. In addition, the presence of a two-dimensional electron gas containing massive electrons (parabolic dispersion) due to charge accumulation at the surfaces could lead to another two-dimensional transport channel \cite{Bianchi10,Brahlek15,Mooshammer18}.

\subsection{Nanoribbons}

\begin{figure}[htb]
	\centering
\includegraphics[width=0.95\columnwidth]{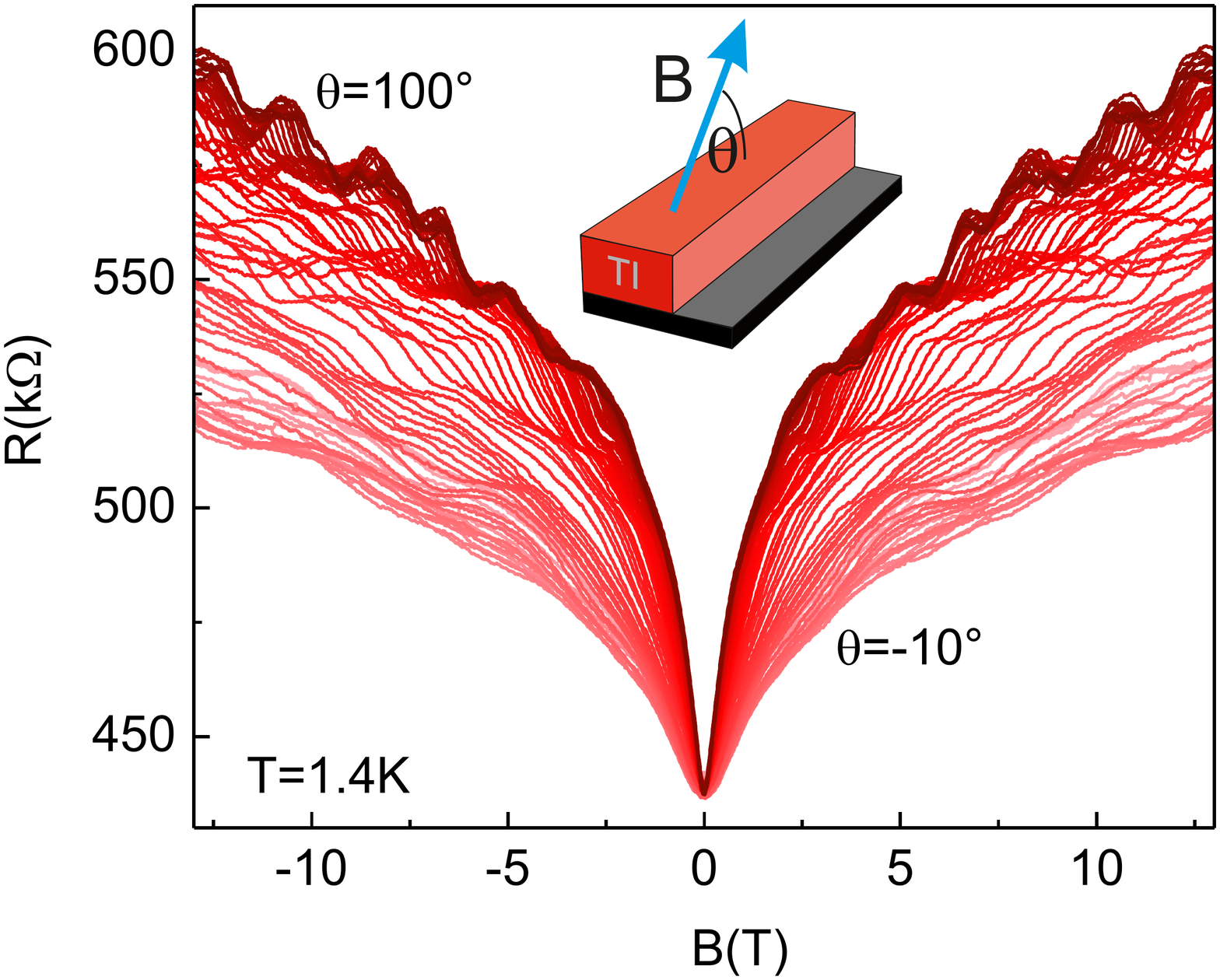}
	\caption{Magnetoresistance of a 10-$\,\mathrm{\mu m}$-long and 50-$\,\mathrm{nm}$-wide nanoribbon at different tilt angles of the magnetic field. The measurements were performed at 1.4\,K. The measurements from $\theta=-10^\circ$ to $\theta=10^\circ$ were done at a stepping of $1^\circ$, from $\theta=10^\circ$ to $\theta=100^\circ$ a stepping of $2^\circ$ was chosen. The inset shows the orientation of the magnetic field $B$ with respect to the upper surface of the ribbon. }
	\label{fig:TI-ribbon-R}
\end{figure}

We now turn to the transport experiments on a selectively-grown nanoribbon with a length of $10\,\mathrm{\mu m}$, a width of $50\,\mathrm{nm}$, and a thickness of $29\,\mathrm{nm}$. The measurements were performed in a two-terminal configuration at a temperature of 1.4\,K. In Fig.~\ref{fig:TI-ribbon-R} the magnetoresistance is shown as a function of tilt angle $\theta$ of the magnetic field. The sample shows a resistance of about $440\,\mathrm{k\Omega}$ at $0\,\mathrm{T}$, corresponding to a resistivity of $\rho= 6.4 \times 10^{-3} \,\Omega\mathrm{cm}$. We attribute the higher resistivity compared to values obtained from the Hall bar measurements to the additional boundary scattering contribution.  

\begin{figure}[htb]
	\centering
\includegraphics[width=0.99\columnwidth]{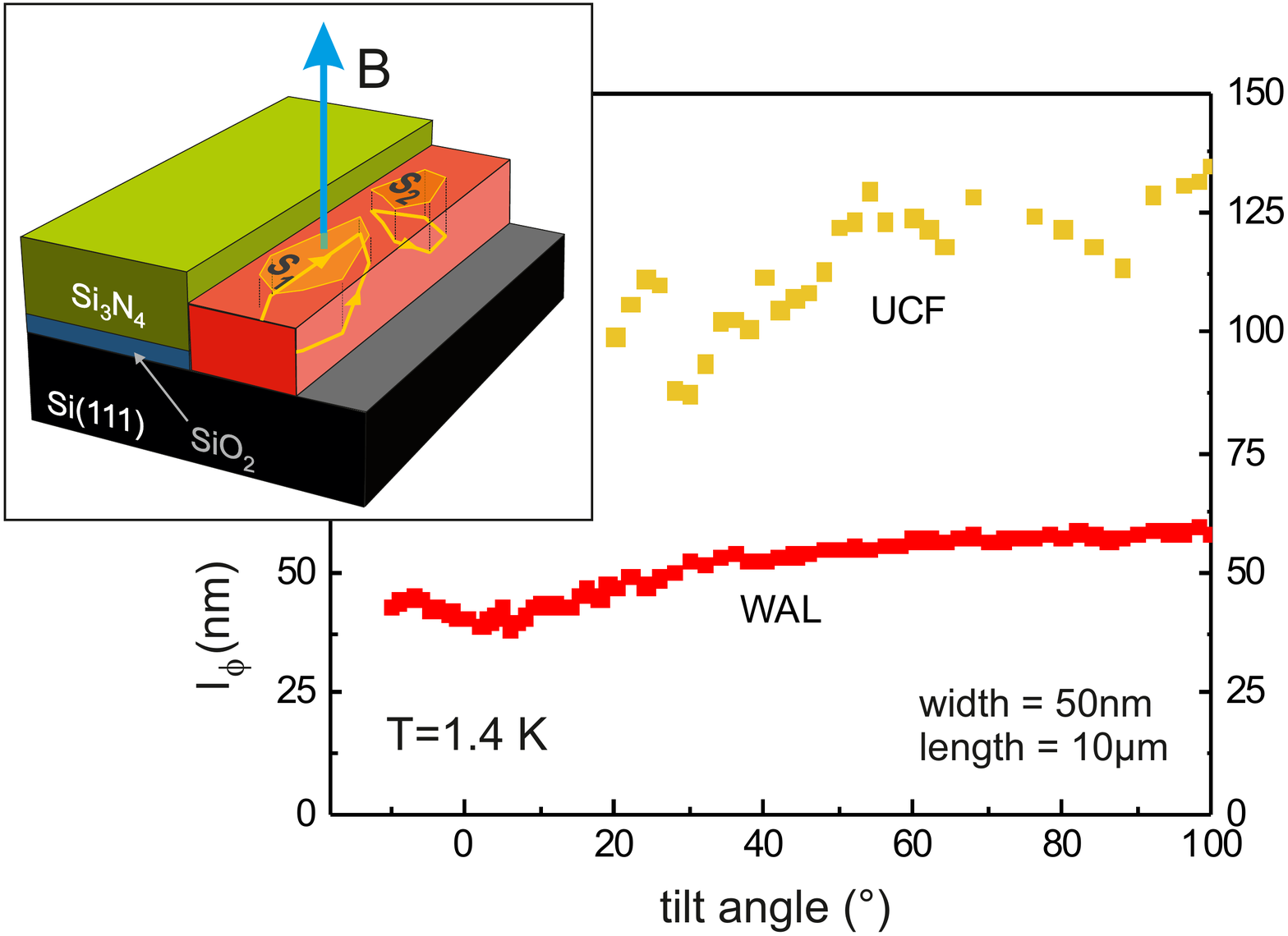}
 	\caption{Phase-coherence length in the nanoribbon as a function of tilt angle of the magnetic field for WAL by fitting the experimental data shown in Fig.~\ref{fig:TI-ribbon-R} according to Eq.~\eqref{eq:DeltaR} (red symbols) and from UCF theory (yellow symbols) applied to the same data. Inset: Schematic arrangement of the nanoribbon with two exemplary closed loops and their projected enclosed areas $S_1$ and $S_2$ with respect to a perpendicular magnetic field $B$.}
	\label{fig:Lphi-WAL-ribbon-angle}
\end{figure}

As found in the measurements on Hall bar structures, the curves possess a dip feature at zero magnetic field due to the weak antilocalization effect. In addition, one finds reproducible resistance modulations, which will be addressed in detail below. Since the phase-coherence length $l_\phi$ extracted from the Hall bar measurements exceeds the ribbon width we applied a model for quasi 1-dimensional structures to fit the experimental data and to extract $l_\phi$ \cite{Altshuler81a,Beenakker88,Kurdak92}. For dirty metals in the strong spin-orbit scattering limit the correction of the resistance due to weak antilocalization can be expressed by a quasiclassical approach \cite{Liang10} 
\begin{equation}
\Delta R = -\frac{1}{2}\frac{R^2}{L} \frac {e^2}{\pi \hbar}
\left( \frac{1}{l_\phi(\theta)^2}+ \frac{1}{l_{B,\perp}(\theta)^2} + \frac{1}{l_{B,||} (\theta)^2}
\right)^{-1/2} \; .\label{eq:DeltaR} 
\end{equation}
Here, $l_\phi$ is the phase-coherence length at a tilt angle $\theta$; $w$ and $d$ are the width and thickness of the ribbon, respectively, and $L$ is the contact separation. Furthermore, the magnetic dephasing lengths for the perpendicular and parallel components of the magnetic field are defined as $l_{B,\perp}=\sqrt{3}\hbar/e B \sin(\theta) w$ and $l_{B,|| }=\sqrt{2\pi}\hbar/e B \cos (\theta) \sqrt{w d} $, respectively \cite{Altshuler81a,Beenakker88,Liang10}. A possible contribution of the Zeeman effect was neglected, since it could be shown by weak localization measurements on topological insulator layers that this contribution is not relevant in the low field range \cite{Wang12a,Chen11}.
The model was fitted to the data for the different sample orientations. Fig. \ref{fig:Lphi-WAL-ribbon-angle} shows the results of the fit (red symbols). For a magnetic field aligned perpendicularly to the substrate plane ($\theta=90^\circ$) we find a phase-coherence length of around $60\,\mathrm{nm}$, which is reduced to about $40\,\mathrm{nm}$ for an in-plane magnetic field. Thus, an angular dependence of $l_{\phi}$ is extracted from the WAL which is not observed in isotropic materials \cite{Liang10}. This indicates an anisotropy of the material favouring coherent transport parallel to the substrate surface. The phase-coherence length of around $60\,\mathrm{nm}$ for a perpendicular field is somewhat smaller than the value gained from the Hall bar structure. A possible reason might be the effect of boundary scattering on the phase-coherence length in case of the nanoribbon.  

The modulations of the magnetoresistance $R(B)$ found in Fig.~\ref{fig:TI-ribbon-R} can be attributed to universal conductance fluctuations \cite{Lee85,Altshuler85b}. As mentioned above, these fluctuations originate from the Aharonov--Bohm type interference of closed-loop electron trajectories of different size and orientation (cf. schematic illustration in Fig.~\ref{fig:Lphi-WAL-ribbon-angle}) \cite{Sacksteder18}. By analyzing the correlation field $B_C$ of the conductance fluctuations the characteristic phase-coherence length for the interference in the loops can be estimated \cite{Lee85,Lee87}. The correlation field is determined via the autocorrelation function $F(\Delta B) = \langle \delta G(B + \Delta B) \delta G(B) \rangle$ by $F(B_C) = \frac{1}{2}F(0)$. Here, $\delta G(B)$ are the  conductance fluctuations after subtracting the slowly varying background and $\langle \dots \rangle$ denotes the average over the magnetic field. For a quasi-one-dimensional transport channel in the dirty limit $l_e \ll w$ the relation between $B_C$ and $l_\phi$ is expressed by $B_c=\gamma \, \Phi_0/(l_\phi w)$ \cite{Lee87,Beenakker88a}. For the prefactor $\gamma$ we choose $0.42$ \cite{Beenakker88a} for $l_\phi$ larger than the thermal length $l_T=\sqrt{\hbar \mathcal{D}/k_B T}\approx 7\,\mathrm{nm}$, with $\mathcal{D}$ the diffusion constant calculated from the mobility and the charge carrier concentration. 

In Fig. \ref{fig:Lphi-WAL-ribbon-angle} the phase-coherence length $l_{\phi}$ extracted from $B_C$ is plotted for different tilt angles $\theta$. We restricted the analysis to tilt angles  $\theta \geq 20^\circ$ since for smaller values of $\theta$ too few fluctuations occurred in the measured magnetic field range. Once again we find a tendency that $l_\phi$ is anisotropic, i.e. $l_\phi$ is smaller for small tilt angles, similar to our analysis of the WAL measurements. However, the values of $l_\phi$ obtained from the UCF measurements are larger by a factor of about two compared to the values gained from the WAL measurements. The significant difference between the two $l_\phi$ values might have several reasons. First, the $l_\phi$ values evaluated from UCFs originate from measurements in large magnetic fields, in contrast to WAL where $l_\phi$ is determined close to zero magnetic field. Consequently, spin-related scattering mechanisms might yield differently strong contributions in both phenomena. Second, there might be uncertainties in the theoretical analysis of both interference phenomena, e.g. the characteristic $\gamma$ factor in the UCF analysis might deviate from the theoretically predicted value \cite{Bloemers11}. Third, the phase coherence length $l_\phi$ is in the order of the circumference of the nanoribbon. In that case the interference effects in the magnetoconductance are also limited by the circumference in addition to $l_\phi$ \cite{Sacksteder14}.      

\begin{figure}[htb]
	\centering
\includegraphics[width=0.90\columnwidth]{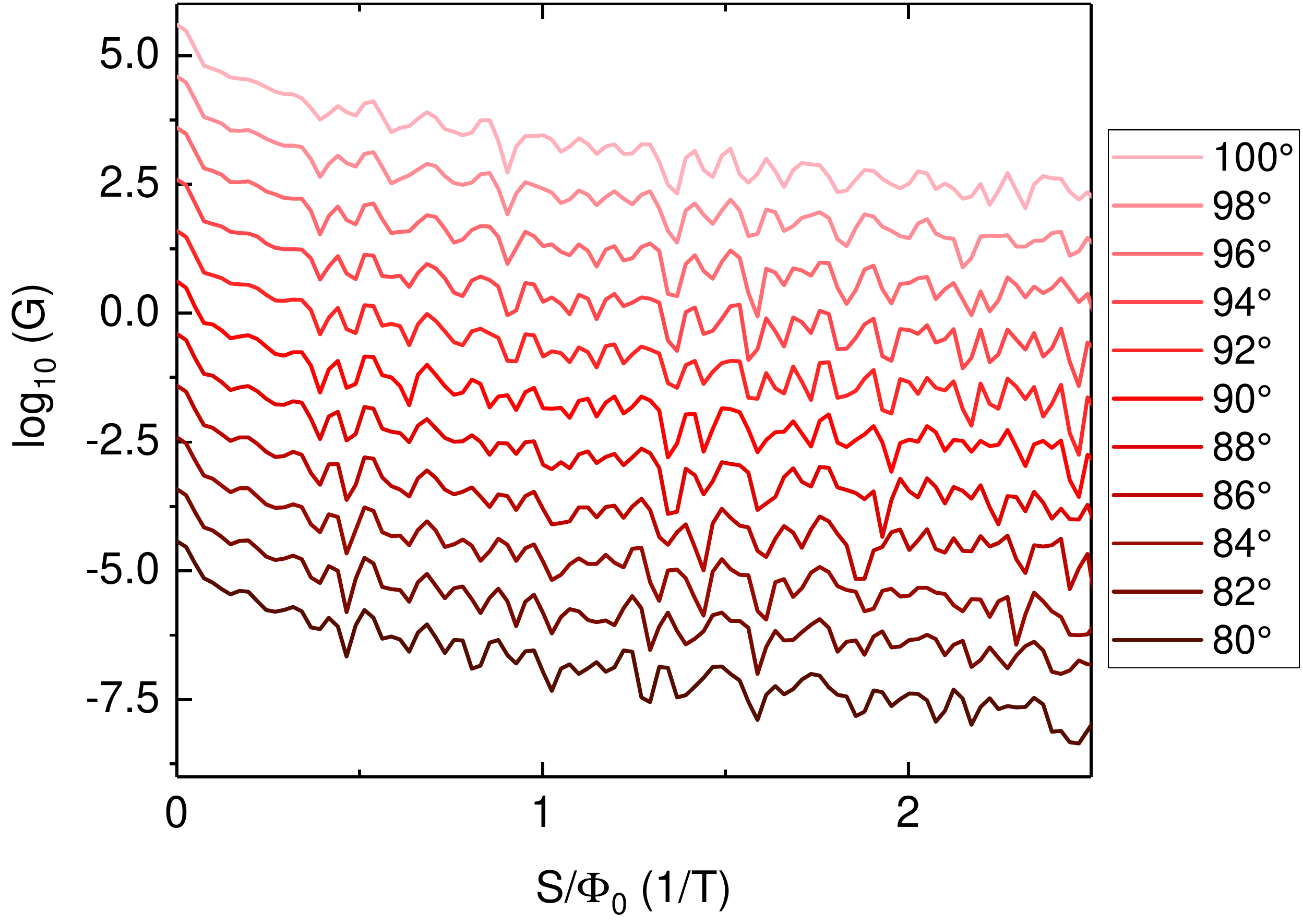}
	\caption{Logarithm of the Fourier amplitude $\log (G)$ vs. $S/\Phi_0$ taken from the magnetoconductance of the nanoribbon at magnetic field tilt angles between 80$^\circ$ and 100$^\circ$, with $S$ the loop cross sectional area and $\Phi_0$ the magnetic flux quantum. The data is plotted with an offset for better visibility. The bottom curve is without offset.}
	\label{fig:FFT_at_different_angles}
\end{figure}

For the universal conduction fluctuations each individual loop contributes to the magnetoconductance with a characteristic period and thus with a particular frequency. Therefore, by analyzing the frequency spectrum of the magnetoconductance detailed information on the loop sizes, i.e. cross sections $\vec{S}_\Lambda$ with respect to an applied magnetic field $\vec{B}$, can be obtained. In order to do so, we calculated the Fourier transform of the magnetoconductance $G(B)=1/R(B)$:
\begin{equation}
G(S/\Phi_0)= \int dB G(B) \exp(-2\pi i B S /\Phi_0) \; ,      
\end{equation}
with $S$ the loop cross sectional area perpendicular to a given orientation of an applied magnetic field. Figure~\ref{fig:FFT_at_different_angles} shows $\log_{10} (G)$, i.e. the logarithm of the Fourier amplitude, as a function of $S/\Phi_0$ ("frequency") in units of $1/T$ for tilt angles between  80$^\circ$ and 100$^\circ$. 
No background was subtracted from the original signal. It is clear that the Fourier spectrum fluctuates indicating that it contains a large number of different frequencies. The spectrum reflects the contribution of different loops with specific cross sectional areas $S$ to the magnetoconductance. The spectrum is reproducible, and changes gradually when the  magnetic field orientation is varied. These changes occur because as the tilt angle changes, each loop's cross-sectional area perpendicular to the magnetic field vector also changes.   

\begin{figure*}[htb]
	\centering
\includegraphics[width=1.55\columnwidth]{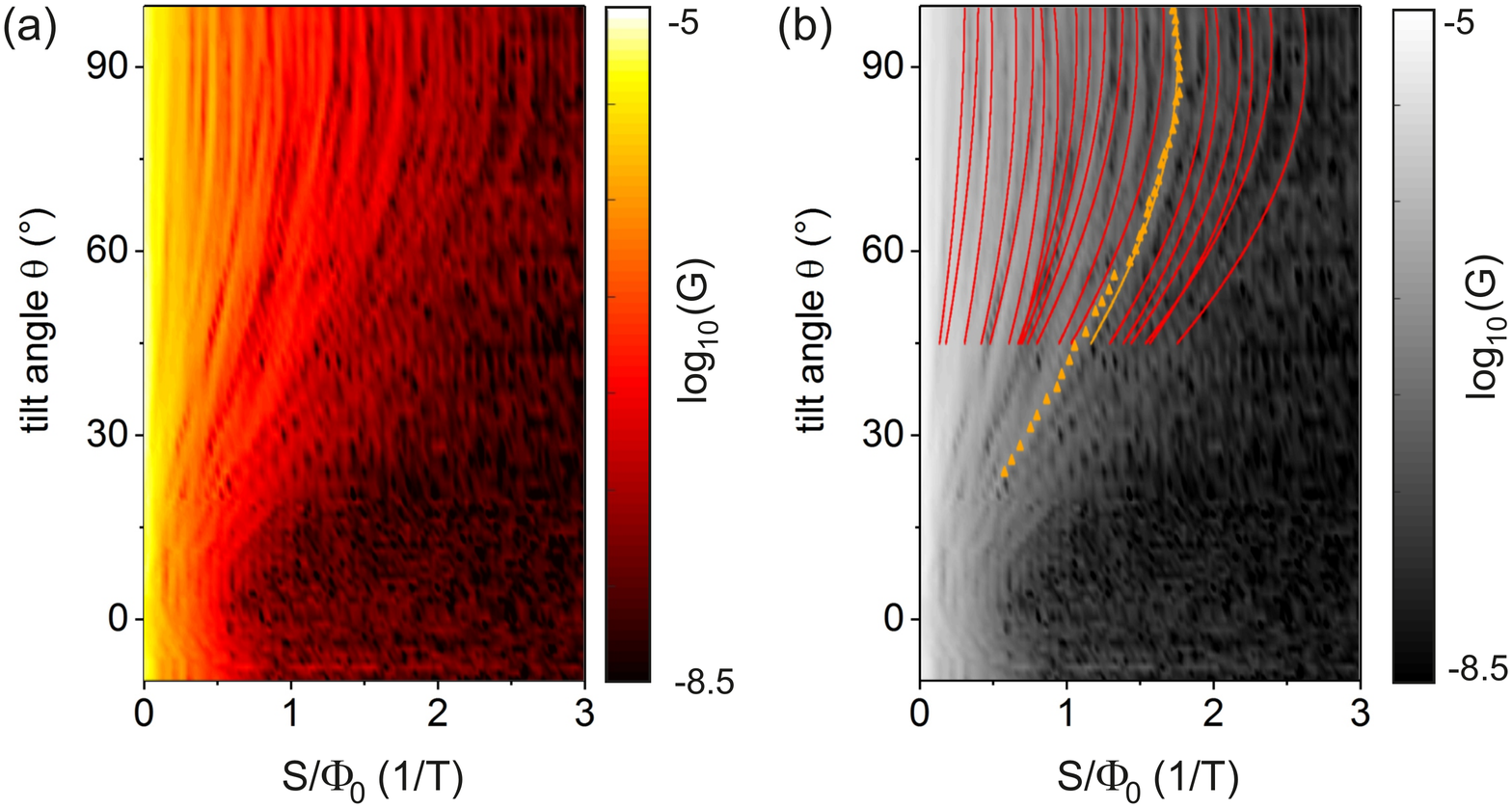}
	\caption{(a) Color plot of the logarithm of the Fourier amplitude $\log(G)$ of the magnetoconductance of the nanoribbon as a function of $S/\Phi_0$ and the magnetic field tilt angle. (b) Corresponding gray-scale plot with a exemplary sequence of dots tracing peaks in the spectrum while the tilt angle changes. The red lines are fits to different dot sequences according to Eq.~(\ref{eq:projection}).}
	\label{fig:FFT-ribbon-angle}
\end{figure*}

Collecting the traces corresponding to different magnetic field orientations in a color plot confirms that there are systematic patterns in the fluctuations of the Fourier spectrum, as shown in Fig.~\ref{fig:FFT-ribbon-angle}(a). 
One can clearly see that for a  perpendicular magnetic field the oscillation frequency range is large whereas for a parallel magnetic field the range is considerably smaller, i.e. the characteristic brighter stripes are getting closer towards smaller magnetic field tilt angles $\theta$. Since the frequency is directly proportional to the loop cross sectional area $S$ with respect to the magnetic field orientation, one can directly deduce that for the perpendicular case larger-area loops contribute. This is indeed plausible, since for that orientation, the maximum phase-coherent loop area is approximately given by $S=l_\phi w$, with $l_\phi$ the phase-coherence length and $w$ the nanoribbon width, assuming $w < l_\phi$. One finds that at $\theta=90^\circ$ the frequency contributions diminish at about $2.5\,\mathrm{T}^{-1}$ corresponding to an area of approximately $10^4$\,nm$^2$ resulting in a phase-coherence length $l_\phi$ of 200\,nm. This value is close but somewhat larger than the corresponding value determined from the correlation field. For the parallel magnetic field orientation the maximum loop area is given by the cross section $S=w d=1450$\,nm$^2$ of the ribbon, providing that the thickness $d$ of the (Bi$_{0.57}$Sb$_{0.43}$)$_2$Te$_3$ layer is smaller than $l_\phi$. The corresponding value of $S/\Phi_0=0.35\,T^{-1}$ fits very well to the range found in Fig.~\ref{fig:FFT-ribbon-angle}(a) at $\theta=0$. However, we do not find a distinct single Aharonov--Bohm peak in the spectrum at $0.35\,T^{-1}$, as it would be expected if the transport solely takes place in the topologically protected surface states or in the accumulation layer of massive electrons at the interface. One reason might be that the thickness of the nanoribbons varies by one or two quintuple layers, resulting in a variation of the cross-section along the ribbon. We furthermore observe some frequency contributions below $0.35\,T^{-1}$ indicating that smaller size loops in the bulk contribute as well.

In Fig.~\ref{fig:FFT-ribbon-angle}(a) we find a stripe-like pattern of peaks in the Fourier spectrum which is getting closer towards smaller tilt angles. Each stripe can be assigned to a distinct phase-coherent loop \cite{Sacksteder18}. In order to gain more precise information on the loop orientation with respect to the magnetic field we followed each trace and determined a set of corresponding discrete points in the $S/\Phi_0$ -- $\theta$ plane, as indicated by the dots in Fig.~\ref{fig:FFT-ribbon-angle}(b). To each sequence of data points we performed a fit according to 
\begin{equation}
\frac{S_\Lambda (\theta)}{\Phi_0} = \frac{S_{\Lambda,\mathrm{max}}} {\Phi_0} \sin \left(\theta + \gamma_\Lambda \right) \; , \label{eq:projection}
\end{equation}
with $S_\Lambda (\theta)$ and $S_{\Lambda,\mathrm{max}}$ the projected and maximum loop area, respectively, while $\gamma_\Lambda$ is the tilt angle of the maximum loop area with respect to the substrate plane. Some exemplary fits are shown in Fig.~\ref{fig:FFT-ribbon-angle}(b) (red lines). We find that most of the lines correspond to loops with small tilt angles $\gamma_\Lambda$ in the range of $\pm 7^\circ$. In fact, we could not identify sine-like lines with larger angles. Thus, we conclude that the interference loops have to be oriented within the plane parallel to the substrate mainly. Indeed, for the largest observed loop area corresponding to $l_\phi=200$\,nm, we found a maximum tilt angle of $\gamma = \sin^{-1} (d/l_\phi)=8.3^\circ$ based on the nanoribbon thickness $d=29$\,nm. This value fits well to the observed range of angles. From the fact that we find finite tilt angles, we can infer that the phase-coherent loops extend over different quintuple layers, i.e. the weak van der Waals bonding does not prevent an extension of the electron waves across the quintuple layers completely. However, since the tilt angles are only found in a small range, we can nevertheless conclude, that the transport preferentially occurs in a plane parallel to the quintuple layers, i.e. the transport is indeed anisotropic \cite{Weyrich19}. In the supplementary information, we support our conclusions drawn here, by modelling the Fourier spectrum for a typical set of loops and deducing from that the magnetoconductance fluctuations.

\section{Conclusion}

In conclusion, by using selective-area molecular beam epitaxy we succeeded to grow topological insulator Hall bar structures as well as nanoribbons. Low temperature magnetotransport experiments on these structures revealed signatures of weak antilocalization. From these measurements a phase-coherence length in the order of 100\,nm was extracted. By performing measurements in a tilted magnetic field we concluded, that the phase-coherence length for interference in time-reversed closed loops mainly oriented parallel to the quintuple layers is larger than in loops containing path segments perpendicular to the quintuple layers. 

For the nanoribbons we also observed universal conductance fluctuations. By performing a Fourier transform of the fluctuation pattern we were able to identify a series of distinct phase-coherent closed-loop trajectories with areas which can be explained in terms of nanoribbon dimensions and phase-coherence length. In accordance with the weak antilocalization measurements we concluded from measurements at different magnetic field tilt angles that the loops are predominately located parallel to the quintuple layers. Thus, the observed anisotropy favours coherent transport along the quintuple layers hence supports our lateral growth technique for coherent transport applications and devices. We expect that the present approach towards analyzing universal conductance fluctuations at different magnetic field orientations by Fourier transforms will be a powerful tool for better understanding conducting pathways in disordered topological materials and nanostructures thereof. 

In the present samples, we could not avoid a significant bulk conductance contribution due to background doping. In future structures, one should try to minimize the background doping either by using optimized ternary or quaternary topological insulators, by means of heterostructures or by gate control. Latter could also help to avoid the two dimensional contribution of massive electrons from the conduction band in the surface accumulation layer. If that is once achieved it should be possible to observe a single peak in the Fourier spectrum for an in-plane magnetic field resulting from Aharonov--Bohm type oscillations due to phase-coherent transport in the topologically protected surface states.    

\section{Acknowledgment}
\noindent We thank Herbert Kertz for technical assistance. This work was supported by the Virtual Institute for Topological Insulators (VITI), which is funded by the Helmholtz Association.  VES thanks S. Bl\"ugel, the Peter Gr\"unberg Institute, and S. Kettemann for hospitality.

\end{document}